\documentclass[12pt]{article}

\begin{document}

\title{\bf Two-Loop Superstrings in Hyperelliptic Language I:
the Main Results}

\author{Zhu-Jun Zheng\thanks{Supported in part by Math. Tianyuan Fund
with grant Number 10226002 and the Natural Science Foundation of
Educational Committee of Henan Province with grant Number
2000110010.
 }\\ \\
Institute of Mathematics, Henan University \\
Kaifeng 475001, P. R. China\\ and\\
Institute of Theoretical Physics,
Chinese Academy of Sciences\\
P. O. Box 2735,  Beijing 100080, P. R. China \\  \\
Jun-Bao Wu \\School of Physics, Peking University \\
Beijing 100871, P. R. China\\ \\
Chuan-Jie Zhu\thanks{Supported in
part by fund from the National Natural Science Foundation of China
with grant Number
90103004.} \\
Institute of Theoretical Physics,
Chinese Academy of Sciences\\
P. O. Box 2735,  Beijing 100080, P. R. China}

\maketitle
\newpage

\begin{abstract}
Following the new gauging fixing method of D'Hoker and Phong, we
study two-loop superstrings in hyperelliptic language. By using
hyperelliptic representation of genus 2 Riemann surface we derive
a set of identities involving the Szeg\"o kernel. These identities
are used to prove the vanishing of the cosmological constant and
the non-renormalization theorem point-wise in moduli space by
doing the summation over all the 10 even spin structures. Modular
invariance is maintained at every stage of the computation
explicitly. The 4-particle amplitude is also computed and an
explicit expression for the chiral integrand is obtained. We use
this result to show that the perturbative correction to the $R^4$
term in type II superstring theories is vanishing at two loops.

In this paper, a summary of the main results is presented with
detailed derivations to be provided in two subsequent
publications.
\end{abstract}


\section{Introduction}

Although we believe that superstring theory is finite in
perturbation at any order \cite{GreenSchwarz1, GreenSchwarz2,
GreenSchwarz3,Martinec}, a rigorous proof is still lacking despite
great advances in the covariant formulation of superstring
perturbation theory \'a la Polyakov. The main problem is the
presence of supermoduli and modular invariance in higher genus. At
two loops these problems were solved explicitly by using the
hyperelliptic formalism in a series of papers
\cite{GavaIengoSotkov, IengoZhu1, Zhu, IengoZhu2, IengoZhu3}. The
explicit result was also used by Iengo \cite{Iengo} to prove the
vanishing of perturbative correction to the $R^4$ term
\cite{GrossWitten} at two loop, in agreement with the indirect
argument of Green and Gutperle \cite{GreenGutperle}, Green,
Gutperle and Vanhove \cite{Green2}, and Green and Sethi
\cite{GreenSethi} that the $R^4$ term does not receive
perturbative contributions beyond one loop. Recently, Stieberger
and Taylor \cite{Stieberger}  also used the result of
\cite{IengoZhu2} to prove the vanishing of the heterotic two-loop
$F^4$ term. For some closely related works we refer the reader to
the reviews \cite{Green3, Kiritsis}. In the general case, there is
no satisfactory solution. For a review of these problem we refer
the reader to \cite{DHokerPhong1, DHokerPhong6}.

Recently two-loop superstring was studied by D'Hoker and Phong. In
a series of papers \cite{DHokerPhong2, DHokerPhong3, DHokerPhong4,
DHokerPhong5} (for a recent review see \cite{DHokerPhong6}),
D'Hoker and Phong found an unambiguous and slice-independent
two-loop superstring measure on moduli space for even spin
structure from first principles.

Although their result is quite explicit, it is still a difficult
problem to use it in actual computation. In \cite{DHokerPhong4},
D'Hoker and Phong used their result to compute explicitly the
chiral measure by choosing the split gauge and proved the
vanishing of the cosmological constant and the non-renormalization
theorem \cite{DHokerPhong7, Martinec}. They also computed the
four-particle amplitude in another forthcoming paper
\cite{DHokerPhong8}. Although the final results are exactly the
expected, their computation is quite difficult to follow because
of the use of theta functions.\footnote{In \cite{Lechtenfeld5},
the two-loop 4-particle amplitude was also computed by using theta
functions. Its relation with the previous explicit result
\cite{IengoZhu2} has not been clarified.}  Also modular invariance
is absurd in their computations because of the complicated
dependence between the 2 insertion points (the insertion points
are also spin structure dependent).

In the old works \cite{GavaIengoSotkov, IengoZhu1, Zhu, IengoZhu2}
on two-loop superstrings, one of the author (with Iengo) used the
hyperelliptic representation to do the explicit computation at two
loops which is quite explicit and modular invariance is manifest
at every stage of the computations. So it is natural to do
computations in this language by using the newly established
result. As we will report in this paper and in more detail in
\cite{AllZhu2, AllZhu3}, everything is quite explicit in
hyperelliptic language although the algebra is a little bit
involved.

By using the hyperelliptic language we derive a set of identities
involving the Szeg\"o kernel (some identities were already derived
in \cite{IengoZhu1, Zhu, IengoZhu2}). These identities are used to
prove the vanishing of the cosmological constant and the
non-renormalization theorem point-wise in moduli space by doing
the summation over all the 10 even spin structures. Modular
invariance is maintained at every stage of the computation
explicitly. The 4-particle amplitude is also computed and an
explicit expression for the integrand is obtained. We use this
result to show that the perturbative correction to the $R^4$ term
in type II superstring theories is vanishing at two loops,
confirming the computation of Iengo \cite{Iengo} and the the
conjecture of Green and Gutperle \cite{GreenGutperle}. We leave
the proof of the equivalence between the new result and the old
result as a problem of the future.

Here we also note that D'Hoker and Phong have also proved that the
cosmological constant and the 1-, 2- and 3-point functions are
zero point-wise in moduli space \cite{DHokerPhong7}. They have
also computed the 4-particle amplitude \cite{DHokerPhong8}. The
agreement of the results from these two different gauge choices
and different methods of computations would be another proof of
the validity of the new supersymmetric gauge fixing method at two
loops.

\section{Genus 2 hyperelliptic Riemann surface}

First we remind that a genus-g Riemann surface, which is the
appropriate world sheet for one and two loops, can be described in
full generality by means of the hyperelliptic
formalism.\footnote{Some early works on two loops computation by
using hyperelliptic representation are \cite{Knizhnik, Morozov,
Morozov1, Morozov2, Lechtenfeld, Bershadsky, Moore} which is by no
means the complete list.}  This is based on a representation of
the surface as two sheet covering of the complex plane described
by the equation:
\begin{equation}
y^2(z) = \prod_{i=1}^{2g+2} ( z- a_i), \label{covering}
\end{equation}
The complex numbers $a_{i}$, $(i=1,\cdots,2g+2)$ are the $2g+2$
branch points, by going around them one passes from one sheet to
the other. For two-loop ($g=2$) three of them represent the moduli
of the genus 2 Riemann surface over which the integration is
performed, while the other three can be arbitrarily fixed. Another
parametrization of the moduli space is given by the period matrix.

At genus 2, by choosing a canonical homology basis of cycles we
have the following list of 10 even spin structures:
\begin{eqnarray}
\delta_1 \sim \left[ \begin{array}{cc} 1 & 1\\ 1 & 1 \end{array}
\right]  \sim (a_1 a_2  a_3|a_4 a_5 a_6),  & & \delta_2 \sim
\left[
\begin{array}{cc} 1 & 1\\ 0 & 0 \end{array} \right] \sim
(a_1a_2a_4|a_3a_5a_6), \nonumber
\\
\delta_3 \sim \left[ \begin{array}{cc} 1 & 0\\ 0 & 0 \end{array}
\right]  \sim (a_1a_2a_5|a_3a_4a_6),  & & \delta_4 \sim \left[
\begin{array}{cc} 1 & 0\\ 0 & 1 \end{array} \right] \sim
(a_1 a_2 a_6|a_3 a_4 a_5), \nonumber
\\
\delta_5 \sim \left[ \begin{array}{cc} 0 & 1\\ 0 & 0 \end{array}
\right]  \sim (a_1 a_3 a_4|a_2 a_5 a_6), & & \delta_6 \sim \left[
\begin{array}{cc} 0 & 0\\ 0 & 0 \end{array} \right] \sim
(a_1 a_3 a_5|a_2 a_4 a_6), \nonumber
\\
\delta_7 \sim \left[ \begin{array}{cc} 0 & 0\\ 0 & 1 \end{array}
\right]  \sim (a_1 a_3 a_6|a_2 a_4 a_5), & & \delta_8 \sim \left[
\begin{array}{cc} 0 & 0\\ 1 & 1 \end{array} \right] \sim
(a_1 a_4 a_5|a_2 a_3 a_6), \nonumber
\\
\delta_9 \sim \left[ \begin{array}{cc} 0 & 0\\ 1 & 0 \end{array}
\right]  \sim (a_1 a_4 a_6|a_2 a_3 a_5), & & \delta_{10} \sim
\left[
\begin{array}{cc} 0 & 1\\ 1 & 0 \end{array} \right] \sim
(a_1 a_5 a_6|a_2 a_3 a_4). \nonumber
\end{eqnarray}
We will denote an even spin structure as $(A_1 A_2 A_3|B_1 B_2
B_3)$. By convention $A_1= a_1$. For each even spin structure we
have a spin structure dependent factor from determinants which is
given as follows \cite{GavaIengoSotkov}:
\begin{equation}
Q_\delta = \prod_{i <j} (A_i-A_j)(B_i-B_j).
\end{equation}
This is a degree 6 homogeneous polynomials in $a_i$.

At two loops there are two odd supermoduli and this gives two
insertions of supercurrent  at two different points $x_1$ and
$x_2$.  Previously the chiral measure was derived in
\cite{Verlinde,DHokerPhong1} by a simple projection from the
supermoduli space to the even moduli space. This projection does't
preserve supersymmetry and there is a residual dependence on the
two insertion points. This formalism was used in
\cite{GavaIengoSotkov, IengoZhu1, Zhu, IengoZhu2}. In these papers
we found that it is quite convenient to choose these two insertion
points as the two zeroes of a holomorphic abelian differential
which are moduli independent points on the Riemann surface. In
hyperelliptic language these two points are the same points on the
upper and lower sheet of the surface. We denote these two points
as $x_1=x+$ (on the upper sheet) and $x_2=x-$ (on the lower
sheet). We will make these convenient choices again in this paper
and \cite{AllZhu2, AllZhu3}.

\section{Some conventions and useful formulas}

In what follows we will give some formulas which will be used
later. First all the relevant correlators are given by\footnote{We
follow closely the notation of \cite{DHokerPhong3}.}
\begin{eqnarray}
\langle \psi^\mu   (z) \psi^\nu   (w) \rangle & = &
-\delta^{\mu\nu} G_{1/2}[\delta] (z,w) = - \delta^{\mu\nu}S_\delta
(z,w),
\nonumber \\
\langle \partial_z X^\mu  (z) \partial_w X^\nu  (w) \rangle & = &
-\delta^{\mu\nu}\partial_z
\partial_w \ln E(z,w),
\nonumber \\
\langle b(z) c(w) \rangle &=& +G_2 (z,w), \nonumber \\
\langle \beta (z) \gamma (w) \rangle &=& -G_{3/2}[\delta] (z,w),
\end{eqnarray}
where
\begin{eqnarray}
& & S_{\delta}(z,w) = { 1\over z-w} \, { u(z) + u(w) \over 2
\sqrt{u(z) u(w) } } , \\
& & u(z) = \prod_{i=1}^3 \left( z-A_i \over z-B_i\right)^{1/2}, \\
& & G_2(z,w) = -H(w,z) + \sum_{a=1}^3 H(w,p_a) \, \varpi_a(z,z), \\
& & H(w,z) = { 1\over 2(w- z)} \,\left( 1 + { y(w) \over
y(z) }\right) \, { y(w) \over y(z) }, \\
& & G_{3/2}[\delta](z,w) = - P(w,z) + P(w,q_1) \psi_1^*(z) +
P(w,q_2)\psi_2^*(z), \label{eq51} \\
& & P(w,z) = {1\over \Omega(w)}\, S_{\delta}(w,z)\Omega(z),
\end{eqnarray}
where $\Omega(z)$ is an abelian differential satisfying
$\Omega(q_{1,2}) \neq 0$. These correlators were adapted from
\cite{Iengo2}. $\varpi_a(z,w)$ are defined in \cite{DHokerPhong2}
and $\psi^*_{1,2}(z)$ are the two holomorphic $3\over
2$-differentials. When no confusion is possible, the dependence on
the spin structure $[\delta]$ will not be exhibited.

In order take the limit of $x_{1,2}\to q_{1,2}$ we need the following
expansions:
\begin{eqnarray}
G_{3/2} (x_2, x_1) &=& {1 \over x_1 - q_1} \psi ^* _1 (x_2)
      - \psi ^* _1 (x_2) f_{3/2} ^{(1)} (x_2) +O(x_1 - q_1),
\\
G_{3/2} (x_1, x_2) &=& {1 \over x_2 - q_2} \psi ^* _2 (x_1)
      - \psi ^* _2 (x_1) f_{3/2} ^{(2)} (x_1)  +O(x_2 - q_2),
\end{eqnarray}
for $x_{1,2}  \to q_{1,2}$. By using the explicit expression of
$G_{3/2}$ in (\ref{eq51}) we have
\begin{eqnarray}
f_{3/2} ^{(1)} (q_2) & = & - {\partial_{q_2} S(q_1,q_2) \over
S(q_1,q_2)
} + \partial\psi^*_2(q_2), \label{eq54}\\
f_{3/2} ^{(2)} (q_1) & = &   {\partial_{q_1} S(q_2,q_1) \over
S(q_1,q_2) } + \partial\psi^*_1(q_1) = f_{3/2} ^{(1)}(q_2)|_{ q_1
\leftrightarrow q_2 } . \label{eq55}
\end{eqnarray}

The quantity $\psi^*_\alpha (z)$'s are holomorphic $3\over
2$-differentials and are constructed as follows:
\begin{equation}
\psi^*_\alpha (z) = (z-q_\alpha)S(z,q_\alpha)  \,
{y(q_\alpha)\over y(z)} \, , \qquad \alpha = 1, 2.
\end{equation}
For $z=q_{1,2}$ we have
\begin{eqnarray} &  & \psi^*_\alpha (q_\beta ) =
\delta_{\alpha,\beta}, \\
& & \partial \psi^*_1 (q_2) = -\partial \psi^*_2 (q_1) =
S(q_1,q_2) = {i\over 4}S_1(q), \\
& & \partial \psi^*_1 (q_1) =  \partial \psi^*_2 (q_2) =
- {1\over 2} \Delta_1(q),  \\
& & \partial^2  \psi^*_1 (q_1) =  \partial^2 \psi^*_2 (q_2) =
{1\over 16}S_1^2(q)  + {1\over 4}\Delta_1^2(q) + {1\over
2}\Delta_2(q),
\end{eqnarray}
where
\begin{eqnarray}
\Delta_n(x) & \equiv & \sum_{i=1}^6 {
1\over (x - a_i)^n }, \\
S_n(x) & \equiv &   \sum_{i=1}^3 \left[ { 1\over (x - A_i)^n } - {
1\over (x - B_i)^n }\right],
\end{eqnarray}
for $  n = 1, 2$. This shows that $\partial\psi^*_\alpha
(q_{\alpha+1})$ and $\partial^2\psi^*_\alpha(q_\alpha)$ are spin
structure dependent.

Other explicit formulas for $\partial_z \partial_w \ln E(z,w)$
will be given in \cite{AllZhu3}.

\section{The chiral measure: the result of D'Hoker and Phong}

The chiral measure obtained in \cite{DHokerPhong2, DHokerPhong3,
DHokerPhong4, DHokerPhong5} after making the choice $x_\alpha =
q_\alpha$ ($\alpha= 1, 2$) is
\begin{eqnarray}
{\cal A} [\delta] & = & i {\cal Z} \biggl \{ 1  + {\cal X}_1 + {\cal
X}_2 + {\cal X}_3 + {\cal X}_4 +  {\cal X}_5 + {\cal X}_6 \biggr
\},
\nonumber \\
{\cal Z} & = & {\langle  \prod _a b(p_a) \prod _\alpha \delta (\beta
(q_\alpha)) \rangle \over \det \omega _I \omega _J (p_a) } ,
\end{eqnarray}
and the ${\cal X}_i$ are given by:
\begin{eqnarray}
{\cal X}_1 + {\cal X}_6 &=& {\zeta ^1 \zeta ^2 \over 16 \pi ^2}
\biggl [ -\langle \psi(q_1)\cdot \partial X(q_1) \, \psi(q_2)\cdot
\partial X(q_2) \rangle  \nonumber  \\
&& \hskip -1cm
 - \partial_{q_1} G_2 (q_1,q_2) \partial\psi^*_1 (q_2)
 + \partial_{q_2} G_2 (q_2,q_1) \partial\psi^*_2 (q_1)
\nonumber \\
&& \hskip -1cm + 2   G_2 (q_1,q_2) \partial\psi^*_1 (q_2)  f_{3/2}
^{(1)} (q_2) - 2   G_2 (q_2,q_1) \partial\psi^*_2 (q_1)  f_{3/2}
^{(2)} (q_1) \biggr ] \, ,
 \\
{\cal X}_2 + {\cal X}_3 &=&  {\zeta ^1 \zeta ^2 \over 8 \pi ^2}
S_\delta (q_1,q_2) \nonumber \\
&& \hskip  1cm  \times \sum_{a=1}^3 \tilde\varpi_a  (q_1, q_2)
\biggl [ \langle T(\tilde p_a)\rangle + \tilde B_2(\tilde p_a) +
\tilde B_{3/2}(\tilde p_a) \biggr ]\, , \label{eq65}  \\
{\cal X}_4 &=& {\zeta ^1 \zeta ^2 \over 8 \pi ^2} S_\delta
(q_1,q_2) \sum _{a=1}^3 \biggl [ \partial_{p_a} \partial_{q_1} \ln
E(p_a,q_1) \varpi^*_a(q_2) \nonumber \\
& & \hskip  1cm+ \partial_{p_a}
\partial_{q_2} \ln E(p_a,q_2) \varpi ^*_a(q_1) \biggr ]\, ,
 \\
{\cal X}_5 &=& {\zeta ^1 \zeta ^2 \over 16 \pi ^2} \sum_{a=1}^3
\biggl
[ S_\delta (p_a, q_1) \partial_{p_a} S_\delta (p_a,q_2) \nonumber \\
& & \hskip  1cm - S_\delta (p_a, q_2) \partial_{p_a} S_\delta
(p_a,q_1) \biggr ] \varpi_a (q_1,q_2) \, .
\end{eqnarray}
Furthermore, $\tilde B_2$ and $\tilde B_{3/2}$ are given by
\begin{eqnarray}
\tilde B_2(w) & = & -2 \sum _{a=1}^3 \partial_{p_a} \partial_w \ln
E(p_a,w) \varpi^*_a (w) \, , \\
\tilde B_{3/2}(w) &=& \sum_{\alpha=1}^2  \biggr(G_2 (w,q_\alpha)
\partial_{q_\alpha} \psi^*_\alpha (q_\alpha) + {3 \over 2}
\partial_{q_\alpha}
G_2 (w,q_\alpha) \psi^*_\alpha (q_\alpha) \biggr)  \, .
\end{eqnarray}

In comparing with the results given in \cite{DHokerPhong4}, we
have written ${\cal X}_2$, ${\cal X}_3$ together and we didn't
split $T(w)$ into different contributions. We also note that in
eq. (\ref{eq65}) the three arbitrary points $\tilde p_a$
($a=1,2,3$) can be different from the three insertion points
$p_a$'s of the $b$ ghost field. The symbol $\tilde\varpi_a$ is
obtained from $\varpi_a$ by changing $p_a$'s to $\tilde p_a$'s. In
the following computation we will take the limit of $\tilde p_1
\to q_1$ or $q_2$. In this limit we have
$\tilde\varpi_{2,3}(q_1,q_2) = 0$ and $\tilde\varpi_1(q_1,q_2) =
-1$. This choice greatly simplifies the formulas and also makes
the summation over spin structure doable (see  \cite{AllZhu2,
AllZhu3} for more details).

\section{The chiral measure in hyperelliptic language}

The strategy we will follow is  to isolate all the spin structure
dependent parts first. As we will show in the following, the spin
structure dependent factors are just $S_1(q)$, $S_2(q)$,
$S_1^3(q)$, $S_1(p_a)$ and the Szeg\"o kernel if we also include
the vertex operators. Before we do this we will first write the
chiral measure in hyperelliptic language and take the limit of
$\tilde p_1 \to q_1$. The full computations and the complete
results will be presented in \cite{AllZhu2, AllZhu3}. Here we only
present the singular terms and other terms which depend on the
spin structure.

First we have
\begin{equation}
T_{\beta\gamma}(w)    =   - { 3/2\over (w-q_1)^2} -
{\partial\psi^*_1(q_1)\over w-q_1} -{1\over8}\Delta_1^2(q)  - {
1\over 32}S_1^2(q) + O(w-q_1).
\end{equation}
In this limit, the dependence on the abelian differential
$\Omega(z)$ drops out. These singular terms are cancelled by
similar singular terms in $\tilde B_{3/2}(w)$. By explicit
computation we have:
\begin{eqnarray}
& & \tilde B_{3/2}(w)   =     { 3/2\over (w-q_1)^2} +
{\partial\psi^*_1(q_1)\over w-q_1}   - {1\over 4}\Delta_1^2(q) + {
3\over 4}\Delta_2(q) \nonumber \\
& & \qquad  - \left( {1\over p_1-q} \, { (q-p_2)(q-p_3) \over
(p_1-p_2)(p_1-p_3)} \, \Delta_1(q) + ... \right)
\nonumber \\
& & \qquad - { 3\over 2} \left( {1\over (p_1-q)^2} \,
{(q-p_2)(q-p_3) \over (p_1-p_2)(p_1-p_3)} + ... \right) +
O(w-q_1).
\end{eqnarray}
where $...$ indicates two other terms obtained by cyclicly
permutating $(p_1,p_2,p_3)$. By using the above explicit result we
see that the combined contributions of $T_{\beta\gamma}(w)$ and
$\tilde B_{3/2}(w)$ are non-singular in the limit of $w\to q_1$.
We can then take $\tilde p_1\to q_1$ in ${\cal X}_2 + {\cal X}_3$.
In this limit only $a=1$ contributes to ${\cal X}_2+{\cal X}_3$.
This is because $\tilde\varpi_{2,3}(q_1,q_2) = 0$ and
$\tilde\varpi_1(q_1,q_2) = -1$.

Apart from the factor ${\zeta^1 \zeta^2\over 16 \pi^2}$, we have
the following form of the left part of integrand for the
$n$-particle amplitude (by combining the chiral measure and the
left part of the vertex operators):
\begin{eqnarray} {\cal A}_1 +
{\cal A}_6 & = & - \langle \psi(q_1)\cdot\partial X(q_1)
\psi(q_2)\cdot\partial X(q_2) \prod_{i} V_i \rangle
\nonumber \\
& & - (\partial_{q_1}G_2(q_1,q_2) +
\partial_{q_2}G_2(q_2,q_1) ) S(q_1,q_2) \langle  \prod_{i} V_i
\rangle
\nonumber \\
& &   + 2 ( G_2(q_1,q_2) + G_2(q_2,q_1) ) \nonumber \\
& & \times  (\partial\psi^*_1(q_1) S(q_1,q_2) -
\partial_{q_2}S(q_1,q_2) ) \langle \,  \prod_{i} V_i \rangle ,
\label{eq30}
\\
{\cal A}_2 + {\cal A}_3 & = & -2 S(q_1,q_2) \left\{
 \langle ( T_X(q_1) + T_\psi(q_1) )  \prod_{i} V_i
 \rangle \right.
\nonumber \\
& & + (T_{\beta\gamma}(q_1) + T_{bc}(q_1) + \tilde B_{3/2}(q_1) +
\tilde B_2(q_1) )   \left.  \langle \prod_i
V_i \rangle \right\} , \\
{\cal A}_4 & = & - 2  \sum _{a=1}^3 \biggl [  \partial_{p_a}
\partial_{q_1} \ln
E(p_a,q_1) - \partial_{p_a}
\partial_{q_2} \ln E(p_a,q_2) \biggr ] \nonumber \\
& & \qquad \times \varpi ^*_a(q_1) S(q_1,q_2) \,
 \langle \prod_i V_i \rangle , \\
{\cal A}_5 & = &  \sum_{a=1}^3 {1\over (q_1-p_a)^2}
\varpi_a(q_1,q_2)\, S(p_a+,p_a-)  \langle \prod_i V_i \rangle .
\label{eq33}
\end{eqnarray}
In this paper and in \cite{AllZhu2, AllZhu3}, we consider only the
massless particle from the Neveu-Schwarz sector and the left part
of the vertex operator is
\begin{equation}
V_i(k_i,\epsilon_i;z_i,\bar z_i) = ( \epsilon_i \cdot
\partial X(z_i) + i k_i \cdot \psi(z_i) \epsilon_i\cdot\psi(z_i) ) \,
{\rm e}^{ i k_i\cdot X(z_i,\bar z_i) } .
\end{equation}
Because the vertex operator doesn't contain any ghost fields, all
terms involving ghost fields can be explicit computed which we
have done in the above. For the computation of amplitudes of other
kinds of particles (like fermions), one either resorts to
supersymmetry or can use similar method which was used in
\cite{Sen,Zhu2} to compute the fermionic amplitude.

From the above results we see that all the spin structure
dependent parts (for the cosmological constant) are as follows:
\begin{equation}
c_1 S_1(q) + c_2 S_2(q) + c_3 S_1^3(q)+ \sum_{a=1}^3 d_a S_1(p_a),
\end{equation}
where $c_{1,2,3}$ and $d_a$'s are independent of spin structure.
In computing the $n$-particle amplitude there are more spin
structure factors coming from the correlators of $\psi$. These are
explicitly included in eqs. (\ref{eq30})--(\ref{eq33}).

\section{The vanishing of the cosmological constant and
non-renormalization theorem}

The vanishing of the cosmological constant is proved by using the
following identities:
\begin{eqnarray}
& & \sum_\delta \eta_\delta Q_\delta S_n(x) = 0, \\
& & \sum_\delta \eta_\delta Q_\delta S_1^3(x) = 0,
\end{eqnarray}
for $n=1,2$ and arbitrary $x$.

For the non-renormalization theorem we need more identities. By
modular invariance we can easily prove the following ``vanishing
identities":
\begin{eqnarray}
& & \sum_\delta \eta_\delta Q_\delta\left\{ {u(z_1) \over u(z_2)}
- {u(z_2)\over u(z_1)} \right\}     = 0,  \label{eq38} \\
& & \sum_\delta \eta_\delta Q_\delta\left\{ {u(z_1) u(z_2) \over
u(z_3) u(z_4)}
- {u(z_3) u(z_4)\over u(z_1) u(z_2) } \right\}  = 0, \\
&  & \sum_\delta \eta_\delta Q_\delta\left\{ {u(z_1) \over u(z_2)}
+ {u(z_2)\over u(z_1)} \right\}  \, S_n(x)  = 0, \qquad n = 1, 2, \\
& & \sum_\delta \eta_\delta Q_\delta \left\{ {u(z_1) \over u(z_2)}
- (-1)^n {u(z_2)\over u(z_1)} \right\}(S_1(x))^n = 0, \qquad n =1,
2, 3. \label{eq41}
\end{eqnarray}
These identities can be proved by modular invariance and simple
``power counting".  To prove the vanishing of the 3-particle
amplitude we also need a ``non-vanishing identity". This and other
identities needed in the 4-particle amplitude computations are
summarized as follows:
\begin{eqnarray}
& &  \sum_\delta \eta_\delta Q_\delta \left\{ {u(z_1)  u(z_2)
\over u(z_3)u(z_4)} - (-1)^n {u(z_1)u(z_2)\over u(z_3)u(z_4) }
\right\} (S_m(x))^n \nonumber  \\
& &  \qquad \qquad =  {2 P(a) \prod_{i=1}^2\prod_{j=3}^4 (z_i-z_j)
\prod_{i=1}^4(x-z_i) \over y^2(x) \prod_{i=1}^4 y(z_i) } \times
C_{n,m},
\end{eqnarray}
where
\begin{eqnarray}
C_{1,1} & = & 1,  \label{eq991} \\
C_{2,1} & = & - 2 (\tilde z_1 + \tilde z_2 - \tilde z_3 - \tilde
z_4) ,     \\
C_{1,2} & = & \Delta_1(x)  - \sum_{k=1}^4 \tilde z_k   , \\
C_{3,1} & = & 2 \Delta_2(x) - \Delta_1^2(x) + 2 \Delta_1(x)\,
\sum_{k=1}^4 \tilde z_k \nonumber \\
& &  + 4 \sum_{k<l} \tilde z_k \tilde z_l   - 1 2 ( \tilde z_1  +
\tilde z_2 )(\tilde z_3 + \tilde z_4 )  \, ,  \label{eq992} \\
\tilde z_k & = & { 1\over x - z_k}, \\
 P(a) & = & \prod_{i<j}(a_i-a_j).
\end{eqnarray}
$C_{1,1}$ and $C_{1,2}$ were derived in \cite{IengoZhu2}. We will
not derive these formulas here and refer the reader to
\cite{AllZhu3}. You will find some other interesting identities
also in \cite{AllZhu2}. Although other values of $n,m$ also gives
modular invariant expressions, the results are quite
complex.\footnote{This is partially due to the non-vanishing of
the summation over spin structures when we set $z_1=z_3$ or
$z_1=z_4$, etc.} Fortunately we only need to use the above listed
results.

By using these formulas we have:
\begin{eqnarray}
& & \sum_\delta \eta_\delta Q_\delta
S_{\delta}(x,z_1)S_{\delta}(z_1,z_2)S_{\delta}(z_2,z_3)\partial_x
S_{\delta}(z_3,x) \, S_1(x) \nonumber \\
& & \qquad \qquad    = - {P(a)\over16  y^2(x) } \prod_{i=1}^3
{x-z_i\over y(z_i)} . \label{eq112}
\end{eqnarray}
We note that the above formula is invariant under the interchange
$z_i \leftrightarrow z_j$.

By using this result and other ``vanishing identities" given in
eqs. (\ref{eq38})--(\ref{eq41}), we proved  the vanishing of the
cosmological constant and the non-renormalization theorem at two
loops (see \cite{AllZhu2} for details).

\section{The 4-particle amplitude}

The 4-particle amplitude can also be computed explicitly. The
final result for the chiral integrand is:
\begin{eqnarray}
{\cal A} & = &   K(k_i,\epsilon_i) \langle :( \partial X(q_1) +
X(q_2) ) \cdot ( \partial X(q_1) +
\partial X(q_2) ) : \nonumber \\
& &  \times \prod_{i=1}^4 \hbox{e}^{i k_i \cdot X(z_i, \bar z_i)}
\rangle \prod_{i=1}^4 { q -z_i\over y(z_i) }  \nonumber \\
& = &  {  K(k_i,\epsilon_i) \over   \prod_{i=1}^4 y(z_i) } \,
\prod_{i<j} {\rm exp}\left[ - k_i\cdot k_j \ G(z_i,z_j) \right]
 \nonumber \\
&  & \times ( s (z_1z_2 + z_3 z_4) + t(z_1z_4+ z_2 z_3) + u(z_1z_3
+ z_2 z_4)) , \label{eq777}
\end{eqnarray}
where $ K(k_i,\epsilon_i)$ is the standard kinematic factor
appearing at tree level and one loop computations
\cite{GreenSchwarz1, Zhu, IengoZhu2}. $G(z_i,z_j)$ is the scalar
Green function which is given in terms of the prime form
$E(z_i,z_j)$ as follows:
\begin{equation}
G(z,w)   =  - \ln |E(z,w)|^2 + 2 \pi \, {\rm Im}\int_z^w
\omega_I\,  ({\rm Im}\Omega)^{-1}_{IJ}  \, {\rm Im} \int_z^w
\omega_J .
\end{equation}
Here in eq. (\ref{eq777}) we also included the factor ${\cal Z}$
and used the explicit correlators for $\langle
\partial X(z) \partial X(w)\rangle$ and $\langle \partial X(z)
X(w,\bar w) \rangle$ given in \cite{Knizhnik, Zhu} (see
\cite{AllZhu3} for details). As it is expected, the find result is
independent on the insertion points $q_{1,2}$ and $p_a$'s.

For type II superstring theory the complete integrand is
\begin{eqnarray}
{\cal A} & = & c_{II} \, K(k_i,\epsilon_i) \langle : (\partial
X(q_1) +\partial X(q_2)) \cdot ( \partial X(q_1) +
\partial X(q_2) ) :  \nonumber \\
& & \times : (\bar{\partial} X(\tilde{\bar q}_1) +{\bar\partial}
X(\tilde{\bar q}_2)) \cdot ( \bar{\partial} X(\tilde{\bar q}_1) +
\bar{\partial} X(\tilde{\bar q}_2) ) :  \nonumber \\
& & \times \prod_{i=1}^4 {\rm e}^{i k_i \cdot X(z_i, \bar z_i)}
\rangle \prod_{i=1}^4 { (q -z_i)(\tilde{\bar q}-\bar z_i)
\over |y(z_i)|^2 } \nonumber \\
&  = & c_{II} \, { K(k_i,\epsilon_i) \over 2 \prod_{i=1}^4 |y(z_i)
|^2 } \,
\prod_{i<j} {\rm exp}\left[ - k_i\cdot k_j \ G(z_i,z_j) \right]
\nonumber \\
&  & \times | s (z_1z_2 + z_3 z_4) + t(z_1z_4+ z_2 z_3) + u(z_1z_3
+ z_2 z_4)|^2 , \label{eqr4}
\end{eqnarray}
which is  independent the left-mover insertion points $q_{1,2}$
and also the right part insertion points $\tilde q_{1,2}$.

The amplitude is obtained by integrating over the moduli space. At
two loops, the moduli space can be parametrized either by the
period matrix or three of the six branch points. We have
\begin{eqnarray}
{ A}_{II} & = &  c_{II}\, K(k_i,\epsilon_i) \, \int {
\prod_{i=1}^6 {\rm d}^2 a_i/{\rm d} V_{pr} \over T^5  \,
\prod_{i<j} |a_i - a_j|^2 } \,  \nonumber \\
& & \times \prod_{i=1}^4 {  { \rm d}^2 z_i \over |y(z_i)|^2 } \,
\prod_{i<j} {\rm exp}\left[ - k_i\cdot k_j \ G(z_i,z_j) \right]
\nonumber \\
&  & \times | s (z_1z_2 + z_3 z_4) + t(z_1z_4+ z_2 z_3) + u(z_1z_3
+ z_2 z_4)|^2 , \label{eqr5}
\end{eqnarray}
where ${\rm d} V_{pr} = { {\rm d}^2 a_i {\rm d}^2 a_j {\rm d}^2
a_k \over |a_{ij}a_{jk}a_{ki}|^2}$ is a projective invariant
measure and $c_{II}$ is a constant which should be determined by
factorization or unitarity (of the $S$-matrix).

An immediate application of the above result is to study the
perturbative correction to the $R^4$ term at two loops. In the low
energy limit $k_i \to 0$, the chiral integrand is   0 apart from
the kinematic factor because of the extra factors of $s$, $t$ and
$u$ in eq. (\ref{eqr5}). This confirms the explicit computation of
Iengo \cite{Iengo} by using the old result \cite{IengoZhu2, Zhu},
and it is in agreement with the indirect argument of Green and
Gutperle \cite{GreenGutperle}, Green, Gutperle and Vanhove
\cite{Green2}, and Green and Sethi \cite{GreenSethi}.

The finiteness of the amplitude can also be checked. We refer the
reader to  \cite{AllZhu3} for details.

\section*{Acknowledgments}

Chuan-Jie Zhu would like to thank Roberto Iengo for reading the
paper and comments. He would also like to thank E. D'Hoker and D.
Phong for discussions and Jian-Xin Lu and the hospitality at the
Interdisciplinary Center for Theoretical Study, Physics,
University of Science and Technology of China.


\begin{thebibliography}{[20]}

\bibitem{GreenSchwarz1} M. B. Green and J. H. Schwarz,
``Supersymmetric Dual String Theory III", Nucl. Phys. {\bf B198}
(1982) 441.

\bibitem{GreenSchwarz2} J. H. Schwarz, ``Superstring Theory'',
Phys. Reports {\bf 89} (1982) 223--322.

\bibitem{GreenSchwarz3} M. B. Green, ``Lectures on Superstrings", in
{\it Unified String Theories}, edited by M. Green and D. Gross
(World Scientific, 1986).

\bibitem{Martinec} E. Martinec, ``Nonrenormalization Theorems and
Fermionic  String Finiteness", Phys. Lett. {\bf 171B} (1986)
189--194.

\bibitem{GavaIengoSotkov} E. Gava, R. Iengo and G. Sotkov,
``Modular Invariance and the Two-Loop Vanishing of the
Cosmological Constant'', Phys. Lett. {\bf 207B} (1988) 283--291.

\bibitem{IengoZhu1} R. Iengo and C.-J. Zhu, ``Notes on the
Non-Renormalization Theorem in Superstring Theories",  Phys. Lett.
{\bf  212B} (1988) 309--312.

\bibitem{Zhu} C.-J. Zhu,``Two-Loop Computations in
Superstring Theories", Int. J. Mod. Phys. {\bf A4} (1989)
3877--3906.

\bibitem{IengoZhu2} R. Iengo and C.-J. Zhu, ``Two-Loop Computation of
the Four-Particle Amplitude in Heterotic String Theory", Phys.
Lett. {\bf  212B} (1988) 313--319.

\bibitem{IengoZhu3} R. Iengo and C.-J. Zhu, ``Explicit Modular
Invariant Two-Loop Superstring Amplitude Relevant for $R^4$'', J.
High Energy Phys. {\bf 0006} (1999) 011, hep-th/9905050.

\bibitem{Iengo} R. Iengo, ``Computing the $R^4$ Term
at Two Superstring Loops", J. High Energy Phys. {\bf 0202} (2002)
035, hep-th/0202058.

\bibitem{GrossWitten} D. J. Gross and E. Witten, ``Superstring
Modification of Einstein's Equations'', Nucl. Phys. {\bf B277}
(1986) 1.

\bibitem{GreenGutperle} M. B. Green and M. Gutperle, ``Effects of
D-instantons", Nucl. Phys.  {\bf B498} (1997) 195--227,
hep-th/9701093.

\bibitem{Green2} M. B. Green, M. Gutperle and P. Vanhove, ``One Loop
in Eleven Dimension", Phys. Lett. {\bf 409B} (1997) 277--184.

\bibitem{GreenSethi} M. B. Green and S. Sethi, ``Supersymmetry
Constraints on Type IIB Supergravity",  Phys.Rev. {\bf D59} (1999)
046006, hep-th/9808061.

\bibitem{Stieberger} S.~Stieberger and T.R.~Taylor, ``Non-Abelian
Born-Infeld action and type I - heterotic duality  (I): Heterotic
$F^6$ terms at two loops",  Nucl.Phys. {\bf B647} (2002) 49--68,
hep-th/0207026; ``Non-Abelian Born-Infeld Action and Type I -
Heterotic Duality (II): Nonrenormalization Theorems", Nucl. Phys.
{\bf B648} (2003) 3,  hep-th/0209064.

\bibitem{Green3} M. B. Green, ``Interconnections
Between Type II Superstrings, M Theory and N=4 Supersymmetric
Yang--Mills" , hep-th/9903124.

\bibitem{Kiritsis} E. Kiritsis, ``Duality and Instantons in String
Theory",in {\it Proceedings of the 1999 Spring Workshop on
Superstrings and Related Matters}, eds. B. Greene, J. Louis, K.S.
Narain and S. Randjbar-Daemi (World Scientific, Singapore, 2000),
hep-th/9906018.

\bibitem{DHokerPhong1} E. D'Hoker and D.H. Phong, ``The Geometry of
String Perturbation Theory", Rev. Mod. Phys. {\bf 60} (1988)
917--1065.

\bibitem{DHokerPhong6} E.~D'Hoker and D.H.~Phong, ``Lectures on
Two-Loop Superstrings---Hangzhou, Beijing 2002",  hep-th/0211111.

\bibitem{DHokerPhong2} E.~D'Hoker and D.H.~Phong, ``Two-Loop
Superstrings I, Main Formulas'', Phys. Lett. {\bf B529} (2002)
241--255.

\bibitem{DHokerPhong3} E.~D'Hoker and D.H.~Phong, ``Two-Loop
Superstrings II, The chiral Measure on Moduli Space'', Nucl. Phys.
{\bf B636} (2002) 3--60.

\bibitem{DHokerPhong4} E.~D'Hoker and D.H.~Phong, ``Two-Loop
Superstrings III, Slice Independence and Absence of Ambiguities'',
Nucl. Phys. {\bf B636} (2002) 61--79.

\bibitem{DHokerPhong5} E.~D'Hoker and D.H.~Phong, ``Two-Loop
Superstrings IV, The Cosmological Constant and Modular Forms'',
Nucl. Phys. {\bf B639} (2002) 129--181.

\bibitem{DHokerPhong7} E.~D'Hoker and D.H.~Phong, ``Two-Loop
Superstrings V, Scattering Amplitudes, the 1-, 2- and 3-point
functions'', in preparation.

\bibitem{DHokerPhong8} E.~D'Hoker and D.H.~Phong, ``Two-Loop
Superstrings VI, Scattering Amplitudes: the 4-point function", in
preparation.

\bibitem{Lechtenfeld5} O. Lechtenfeld and A. Parkes, ``On Covariant
Multiloop Superstring Amplitudes", Nucl. Phys. {\bf B332} (1990)
39--82.

\bibitem{AllZhu2} Zhu-Jun Zheng, Jun-Bao Wu and  C.-J. Zhu,
``Two-Loop Superstrings in Hyperelliptic Language II: the
Vanishing of the Cosmological Constant and the Non-Renormalization
Theorem", hep-th/0212198.

\bibitem{AllZhu3} Zhu-Jun Zheng, Jun-Bao Wu and  C.-J. Zhu,
``Two-Loop Superstrings in Hyperelliptic Language III: the
Four-Particle Amplitude", hep-th/0212219.

\bibitem{Knizhnik} V. G. Knizhnik, ``Explicit Expression for the
Two-Loop Measure in the Heterotic String Theory", Phys. Lett. {\bf
196B} (1987) 473--476.

\bibitem{Morozov} A. Morozov, ``Two-Loop Statsum of Superstrings",
Nucl. Phys. {\bf B303} (1988) 343--372; ``On the Two-Loop
Contribution to the Superstring Four-Point Function", Phys. Lett.
{\bf 209B} (1988) 473--476.

\bibitem{Morozov1} A. Morozov and A. Pereomov, ``Partition Functions
in Superstring Theory. The Case of Genus Two", Phys. Lett. {\bf
197B} (1987) 115--118; ``A Note on Many-Loop Calculations for
Superstrings in the NSR Formalism", Int. J. Mod. Phys. {\bf A4}
(1989) 1773--1780.

\bibitem{Morozov2} D. Lebdev and A. Morozov, ``Statistical Sums of
Strings on Hyperelliptic Riemann Surfaces", Nucl. Phys. {\bf B302}
(1988) 163--188.

\bibitem{Lechtenfeld} O. Lechtenfeld and A. Parkes, ``On the Vanishing
of the Genus Two Superstring Vacuum Amplitude", Phys. Lett. {\bf
202B} (1988) 75--80.

\bibitem{Bershadsky} M. A. Bershadsky, ``The Superbranch Point and
Two-Loop Correlation in Heterotic String Theory", Mod. Phys. Lett.
{\bf A3} (1988) 91--108.

\bibitem{Moore} G. Moore and A. Morozov, ``Some Remarks on
Two-Loop Superstring Calculations", Nucl. Phys. {\bf B306} (1988)
387--404.

\bibitem{Verlinde} E. Verlinde and H. Verlinde, ``Multiloop
Calculations in Covariant Superstring Theory", Phys. Lett. {\bf
192B} (1987) 95--102.

\bibitem{Iengo2} M. Bonini and R. Iengo, ``Correlation Functions
and Zero Modes on Higher Genus Riemann Surface", Int. J. Mod.
Phys. {\bf A3} (1988) 841--860.

\bibitem{Sen} J. J. Atick and A. Sen, ``Covariant One-Loop Fermion
Amplitudes in Closed String Theories", Nucl. Phys. {\bf B293}
(1987) 317--347.

\bibitem{Zhu2} C.-J. Zhu, ``Covariant Two-Loop Fermion
Amplitude in Closed Superstring Theories", Nucl. Phys. {\bf B327}
(1989) 744--762.

\end{thebibliography}
\end{document}